# Analysis of Infinite Stiffness Using PID Controller

## Rifat Choudhury[1]

[1](Department of Civil Engineering, United International University, Bangladesh)

***Abstract:***
***Background:*** *The design of Proportional-Integral-Derivative (PID) controller is involved to gain infinite stiffness using active vibration control system and it's a modification of previous work. The PID controller is a control loop feedback mechanism which steadily calculates a fallacy value as the difference between a desired set point and a variable process and applies correction based on proportional, integral and derivative terms. In this paper to achieve infinite stiffness magnetic field has been created by passing current, the input of the electromagnet is the output voltage of the sensor. Depending upon the gap between electromagnet and suspended object vibration is isolated by the attractive force of the electromagnet. The control system is also implemented via analog control circuit and microcontroller to achieve infinite stiffness. When maximum force has been applied in this system there will be no deflection i.e., generalized force and deflection. This thesis paper also presents the application, limitation and scope of this system.*
***Materials and Methods:*** *In our system current is controlled by Analog Circuit (PID). The current moves through the electromagnet which creates magnetic force. If the load increases the proximity sensor senses the distance and delivers signal to the controller circuit. After receiving the signal controller circuit increased current supply to the electromagnet through power amplifier circuit. Magnetic force is expanded then and object is levitated in a stable position. If the load is decreased the proximity sensor senses the distance and delivers signal to the controller circuit. Controller circuit receives the signal and decreases current supply to the electromagnet through power amplifier circuit. Thus, magnetic force is increased and object is levitated in a stable position. For model structure and controller design, some elements are used i.e., power supply, operational amplifier. The primary function of a power supply is to convert electrical energy. In this system an operational amplifier, a 24v dc and a 12 v dc power supply are used. An operational amplifier is used for amplifier the supply power.24 v dc power supply is used for converting 220v ac power source into 24v -15v dc which transfer through the electromagnet coil. This system runs by measuring vertical distance as the levitated plate comes across the range of the sensor. 12v dc power supply is used for controller circuit and displacement sensor.*
***Results:*** *Rosuvastatin 20 mg on every other regimen had equal effect when compared to daily dose regimen of atorvastatin 40 mg &rosuvastatin 20mg. The main motive of the paper is to achieve infinite stiffness, which has been completed by using PID controller. Though this system is inherently unstable system because of the system nonlinearity, it has been done by systematic way. Vibration has also been reduced and controlled by controller (PD) system. Another aim of this paper is to reduce cost using PID controller which has also been done successfully since it is very essential to save energy and money using this cost.*
***Conclusion:*** *This paper reveals the feasibility of vibration control techniques for number of diverse applications. This paper is highly applicable for detecting direction i.e., magnetic levitation system and magnetic bearing etc.*
***Key Word****: Stiffness; Vibration; Control; Amplifier; Circuit.*

---------------------------------------------------------------------------------------------------------------------------------

---------------------------------------------------------------------------------------------------------------------------------

## I.    Introduction

Diabetes is now commonly recognized as a coronary heart disease risk equivalent[1,2,3,4]. This is mainly attributed to the high rates of dyslipidemia among diabetic patients which is believed to be one of the major factors accounting for the high percentage of Vibration is a mechanical circumstance where oscillation occurs about an equilibrium point. Vibration may be desirable i.e., tuning fork's motion, the reed in a woodwind instrument or harmonic, loudspeaker's cone or a mobile phone. In many cases, vibration is unacceptable, making unwanted sound and wasting energy. Such vibrations could be caused by unbalance rotating mass uneven friction, or the meshing of gear teeth. Careful designs usually minimize unwanted vibrations. If no energy is dissipated or lost in friction or other resistance during oscillation, the vibration is named as undammed vibration. It is called damped vibration if any energy is lost in this way. In many systems, the damping amount is so small that it can be ignored for most engineering purposes. Lenzen researched about human response to

---------------------------------------------------------------------------------------------------------------------------------

transient vibrations and invented that damping is a crucial factor of controlling vibrations. [1] Passive vibration isolation system has been researched by using High Negative Stiffness System (HNSS) which combine this system (HNSS) with original one. [2] Takeshi Mizuno and Md. Emdadul Hoque presented an application of zero-power controlled magnetic levitation for active vibration control.[3] Md. Emdadul Hoque developed a three-degree-of-freedom isolated vibration system with zero power controlled magnetic suspension technology using an active micro vibration isolator. [4] Negative Stiffness Mechanism system was researched on vibration control and it showed the behavior of NSM vibration isolation systems approximates that of six-DOF linear spring systems up to about 10 to 20 HZ. [5] Overcoming the gravitational force on an object was done by applying a counteracting magnetic field. [6] Electromagnetic levitation using feedback techniques was used to attain stable levitation of a bar magnet. [7] This paper aims to realize the phenomenon of vibration and its uses.it also aims to design and construct the active vibration control system and Realization of infinite stiffness using PID controller. Finally, this is to use cost effective PID controller for vibration control. The main purpose is to obtain a robust, stable and controlled system by tuning the PID controller. It is important to use PID controller to increase the performance and stability of the system. Fast tuning of perfect PID controller parameter yield high quality solution. Also, Various numerical models have been developed to improve the performance of a solar dish by using the pre-heat method [8, 9]. Improved pyrolysis system can also increase the efficiency of the solar dish Stirling engine [10]. However, if combined heating from solar and charcoal firings is provided, it will be more efficient [11, 12].

## II.     Control System

Control system are of two classes i.e., open loop control systems and closed loop control systems. Control system are of two types' i.e., analog system and digital signal controller.

**Active vibration control system**

       In our system current is controlled by Analog Circuit (PID). The current moves through the electromagnet which creates magnetic force. If the load increases the proximity sensor senses the distance and delivers signal to the controller circuit. After receiving the signal controller circuit increased current supply to the electromagnet through power amplifier circuit. Magnetic force is expanded then and object is levitated in a stable position. If the load is decreased the proximity sensor senses the distance and delivers signal to the controller circuit. Controller circuit receives the signal and decreases current supply to the electromagnet through power amplifier circuit. Thus, magnetic force is increased and object is levitated in a stable position. In this position, if we applied load there will be no deflection, which can achieve infinite stiffness. Figure 1 shows block diagram of active vibration using feedback system.

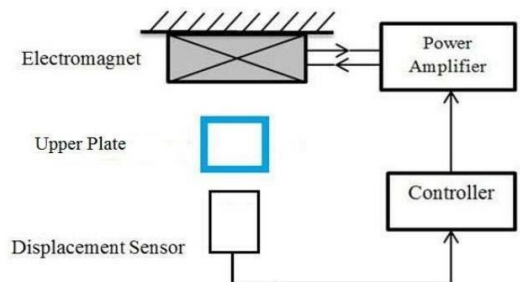

**Fig 1:** Block diagram of active vibration using feedback system

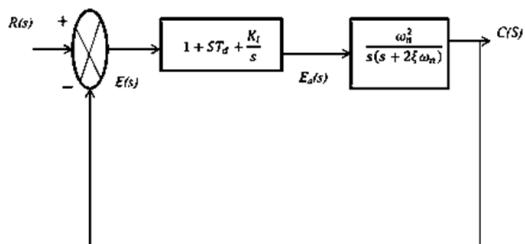

**Fig.2:** PID control

**Proportionalplus Derivative plus Intregal Control (PID)**
In PID control the actuating signal consists of proportional error signal added with derivative and integral of error signal. Figure 2 shows the circuit diagram of PID controller. Equation of PID controller is given here.

$$Ea(s) = E(s) + ST_d E(s) + \frac{K_i}{s} E(s)$$

$$Ea(s) = E(s)\left[1 + ST_d + \frac{K_i}{s}\right]$$

## III. Model Structure

For model structure and controller design, some elements are used i.e., power supply, operational amplifier. The primary function of a power supply is to convert electrical energy. In this system an operational amplifier, a 24v dc and a 12 v dc power supply are used. An operational amplifier is used for amplifier the supply power.24 v dc power supply is used for converting 220v ac power source into 24v -15v dc which transfer through the electromagnet coil. This system runs by measuring vertical distance as the levitated plate comes across the range of the sensor. 12v dc power supply is used for controller circuit and displacement sensor.

The basic model of this system is the PID controller with infinite stiffness. This model structure consists of four stands and there are two rectangular plates. A middle stand is used to hold the sensor and the upper plate hold the electromagnet. There are four springs which are attached with the stands of the model. The diameter of the spring is 1 inch. And the dimension of the total setup is 3 inches height and 6 inches width. The distance of the upper plate and the distance can be varied by using hexagonal nut on the top part of the rectangular plate. The circuit board is attached with the setup.

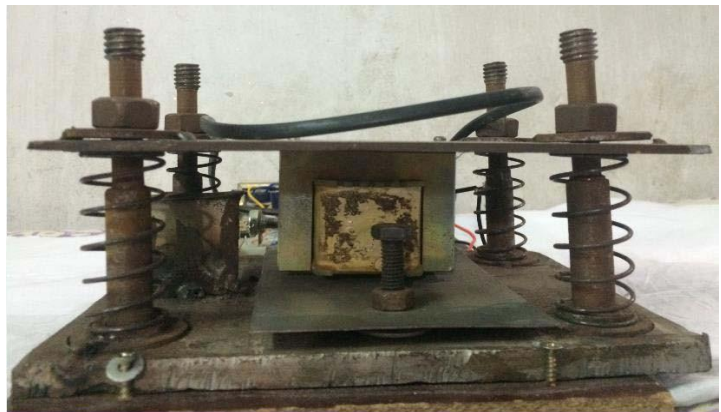

Fig.3: Model structure of vibration control system

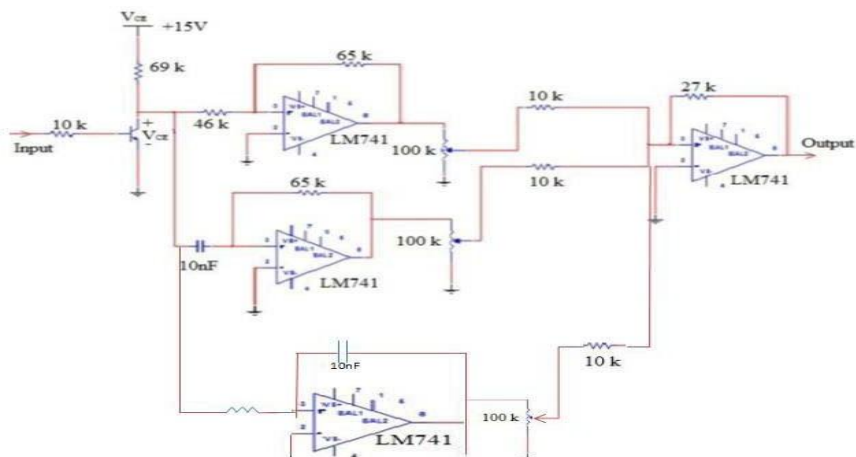

Fig.4: Controller circuit

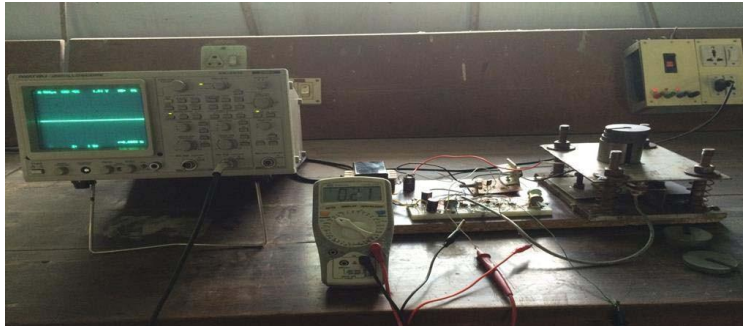

Fig.5: Experimental setup or infinite stiffness

**CONTROLLER DESIGN**

Magnetic Levitation system can be controlled by various ways. In this system, PID controller is used to produce active feedback control system. Here controller is created based on analog circuit with Op-Amp as shown in fig.5. The output is gone through proportional circuit to lift up the object and a derivative circuit to reduce the vibration. After that the output from these circuits is summed in another circuit and combined with an offset power supply. To amplify the output voltage two bi-polar junction transistors (BJT, BD135) are applied to increase the base current of BJT (BC458). So the collector gain of transistor (BC458) is increased and the power input to electromagnet is also amplifier. In this system displacement sensor is used which can detect upto 5mm. Coil, Oscillator, Detection circuit, Output circuit are the components of sensor. Figure 5 displays physical overview of controller circuit. Setup constitute consists of controller circuit board, transformer for circuit board transformer for power amplifier circuit, power amplifier circuit, Steel frame. This setup contains oscilloscope, multimeter and full experimental setup.

## IV. Results

**Sensor Calibration:** The sensitiveness means the sensor senses the distance of the object in a preferable condition. Here the sensor can sense the distance up to 5 mm. From the below graph sensor calibration can be observed. From the fig.6, output shows linear behavior between 2-5 mm. So that the sensor gives value only between these range. Above these distance the sensor become unstable.

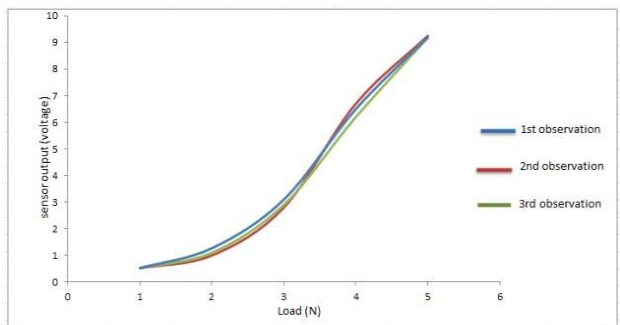

Fig: 6 Sensor calibration (1st 2nd 3rd observations)

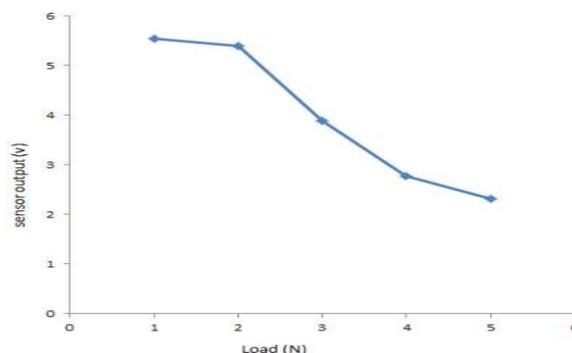

Fig. 7: Variation of sensor output at different load (PD controller)

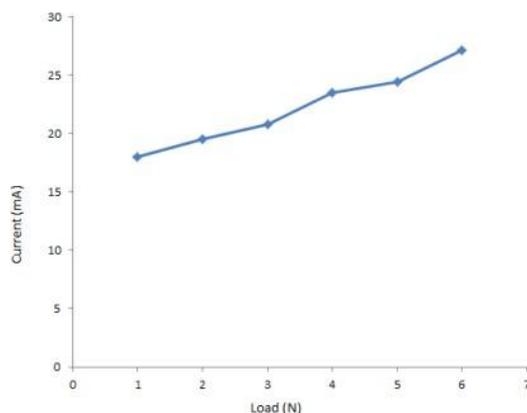

Fig. 8 Variation of current at different load (PD controller)

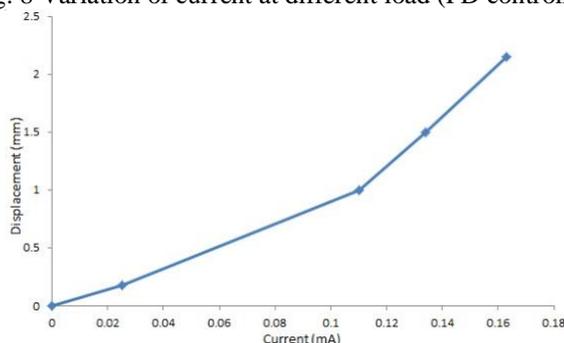

Fig.9 Variation of displacement at different current flow (PD controller

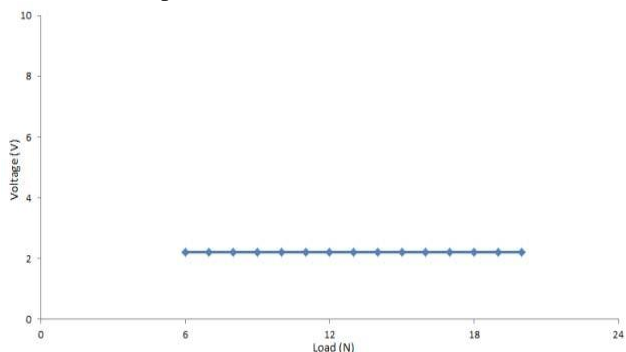

Fig 10: Realization of infinite stiffness (PID)

*A.      Characteristics Curve of Active Vibration Control*
The vibration of the object is further reduced by aligning the object perfectly with the sensor and the electromagnet. Fig.7 shows the increasing load to the sensor output (in volt) which is decreased because the distance between the sensors face and upper vibration isolation plate is decreased. Fig.8 shows the increasing load to controller circuit input (in current) which is increased because the sensor output then decreased. Again, with the increase of load control current is increased to decrease the effect of incoming load.

*B.      Current Co-efficient of Actuator forActive Vibration Control*
Figure 9 represents the different characteristics of the current coefficient. It is clear that with increase of current (for electromagnet) the change of displacement is linear with a certain distance of 1.2mm. After this value the change of displacement is not linear. Considering this displacement and current linearity the current coefficient is determined.

*C.      Realization of infinite stiffness*
Figure 10 represents load (N) vs. sensor output (voltage) graph. Load has been applied at upper plate when PID controller is on; system takes some time to reach stable position and gives some sensor output in voltage.

## V.      Simulations Of Active Vibration Control
To observe the stability of the active vibration control system simulation results are taken in MATLAB software. The value of natural frequency is assumed and it is also assumed that the system is under damped the proportional, derivative and integral gain is obtained. With respect to fixed integral value changes the PD value and output is obtained. Then for different value of PID controller the system output is obtained from the simulations. Considering the linear system, the current coefficient is 7.5. The output of the active vibration control system is shown from figure 11 to 13 for different value of the PD controller.

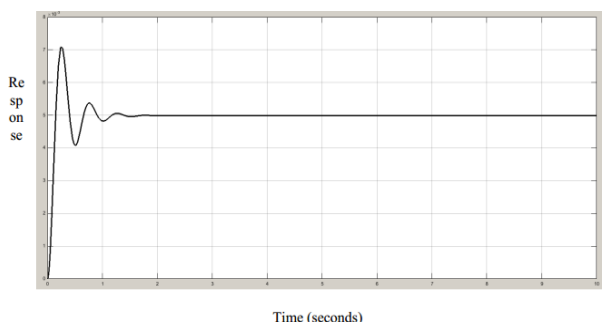

Fig 11: Observation 1 for PD controller

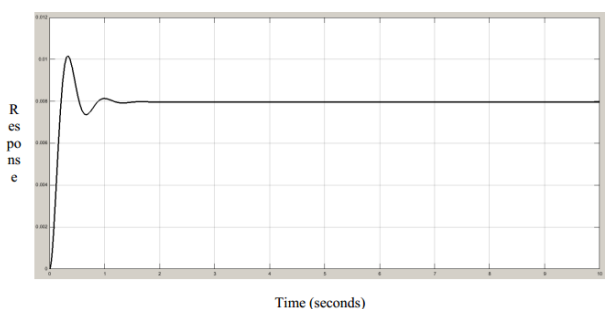

Fig 12: Observation 2 for PD controller

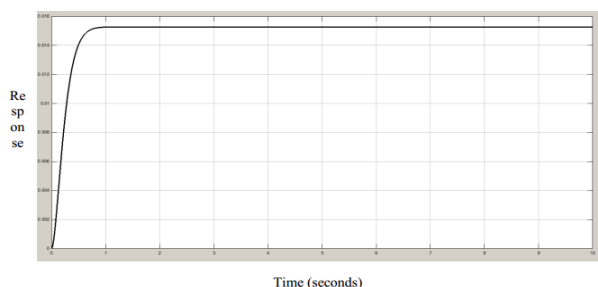

Fig 13: Observation 3 for PD controller

Now the output of the active vibration control system is shown from figure 14 to 17 for different value of the PID controller.

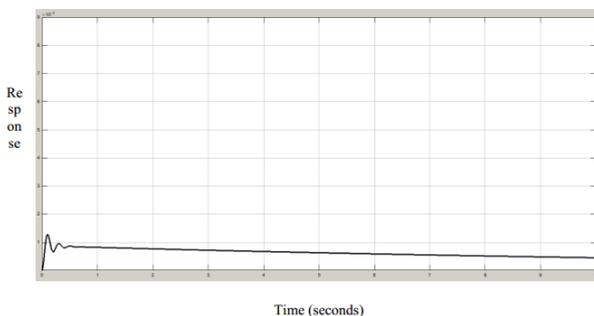

Fig 14: Observation 1 for PID controller

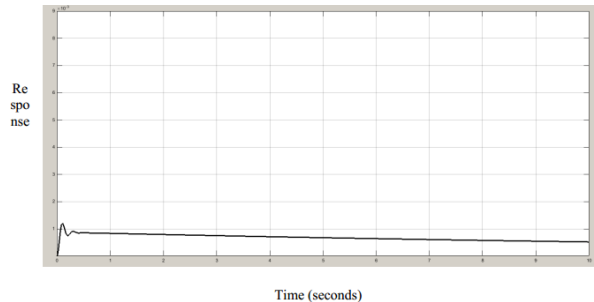

Fig 15: Observation 2 for PID controller

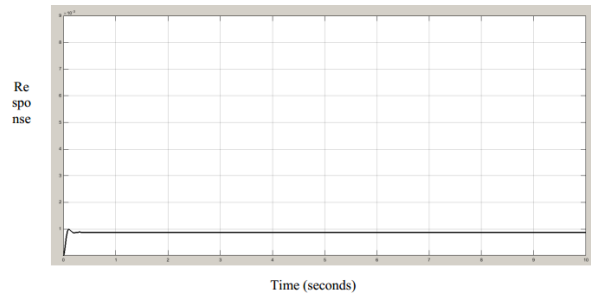

Fig 16: Observation 4 for PID controller

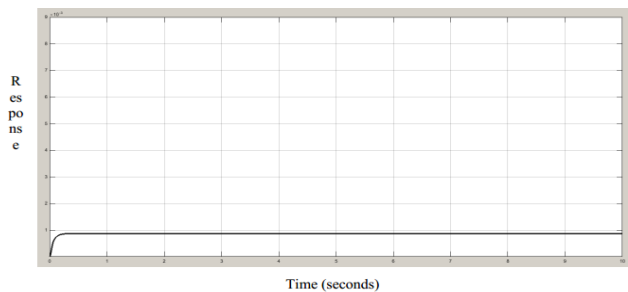

Fig 17: Observation 5 for PID controller

From simulation results it is clear that for greater value of damping coefficient and natural frequency the system quickly reaches in the stable condition and for the lower value of damping coefficient and natural frequency it takes time to reach unstable condition.

## VI.    Conclusion

The main motive of the paper is to achieve infinite stiffness, which has been completed by using PID controller. Though this system is inherently unstable system because of the system nonlinearity, it has been done by systematic way. Vibration has also been reduced and controlled by controller (PD) system. Another aim of this paper is to reduce cost using PID controller which has also been done successfully since it is very essential to save energy and money using this cost. The difficulties involved in developing the sensing the mechanism proved to be the most difficult and time-consuming aspect of this paper. A better understanding of sensing techniques and technologies would have significantly improved the chances of developing a reasonably stable vibration control system. This paper reveals the feasibility of vibration control techniques for number of diverse applications. This paper is highly applicable for detecting direction i.e., magnetic levitation system and magnetic bearing etc.